# 3D Modeling of a Guitar Using a Computer Tomography Scan

Siddique Dalwale, M. Sait Özer, Sebastian Merchel, M. Ercan Altinsoy

*Chair of Acoustics and Haptics TU Dresden, 01062 Helmholtzstraße 18,*
*E-Mail: siddique.dalwale@mailbox.tu-dresden.de*

## Abstract

This paper describes the development of a detailed 3D geometric model of an acoustical guitar. Modeling an instrument is a sophisticated task considering the individual parts and their complex shapes. The geometry of the parts visible from the outside can be measured using appropriate tools, but it is very difficult to measure the details of the internal parts like bracing, heels, and other features by hand through the sound hole. Otherwise, it would be necessary to disassemble the guitar to measure the precise position and dimensions of the parts inside it. Reassembling the guitar could result in improper functioning. To avoid damaging the instrument by disassembling or taking inaccurate measurements through the sound hole, a computer tomography (CT) scan of the guitar body was performed. Using this method, cross-sectional images of the guitar body in all the three dimensions were extracted with 1 mm spacing between adjacent images. In total, approximately 2000 images were generated and used in developing the geometric model of the guitar. The 3D model will be further used to develop a vibro- acoustic simulation model of the guitar.

## Introduction

Around 1.5 million acoustic guitars were sold in the US alone in 2018, according to the estimation of music industries. Production of an acoustic guitar can take months, de-pending on the amount of research and planning that goes into it. To satisfy the needs of various participants, large-scale producers frequently maintain a broad lineup, complicating the management and construction procedures. Additionally, changing models on the manufacturing production line might be expensive. Virtual prototyping has developed into a very helpful technology to produce musical instruments to save time and resources while keeping good product quality [1].

The word guitar comes from the ancient Greek word "Kithara" [2]. It also appears in Egyptian, Arab and Andalusian history. Modern Acoustic guitar was redesigned by Christian Frederick Martin, a German born American Luthier [3]. The body of guitar is made up of the soundboard, the sides, the back, the neck, strings, and several other components [4]. When a string is plucked, the bridge transmits the vibration, which reverberates throughout the top of the guitar. Additionally, these vibrations travel to the back and sides of the guitar and later resonating the air in the cavity of guitar and emitting sound out the sound hole. There have been lots of modifications over the years. For example, stings were made up of sheep intestines in earlier times, later Nylon strings were used and then the steel strings. Cutaways were added to give easy access to lower frets. Body of the guitar has been changing from thinner to thicker. Sound board is made up of wood which is light weight than on the sides and the back. There is a pickguard on the top and it's not only for decoration. Its main purpose is to prevent the guitar top from scratching and striking the top. Bracing has two purposes, one is to pro-vide sufficient strength to support the tension of the strings. Bracing strengthens the top to prevent the neck from warping or part of the body from being lifted when the steel strings are pulled. The second purpose is to create a certain timbre, higher harmonic bars create sharper sounds, while lower ones create softer sounds [5].

To evaluate and study the sound behavior of guitar, it's important to find a proper technique. Experienced guitarist can detect what sort of guitars are the best when they play in real life, but this is not an optimal solution. A simulation method can be used for predicting the Vibro-Acoustic behavior of an instrument at designing process. Therefore, an accurate 3D model is required at the first step. Afterwards, the engineers can use the model to simulate the test cases and optimizing the selected parameters for obtaining desired characteristics.

This study is devoted to create the 3D model of a high end acoustical guitar using reverse engineering techniques and computer added design (CAD) tools.

## Reverse Engineering

It can be challenging to model the geometry of an existing guitar. Although it is simple to measure the geometry of the sections that are visible from the outside, it is not feasible to manually measure the precise bracing features inside the guitar body through the sound hole. Contrary to many technical systems, a guitar cannot simply be disassembled to inspect the position and shape of its internal components without anticipating a different response once the pieces are reassembled. The computed tomography scans (CT) of the guitar body were performed in cooperation with Radiologische Praxis Dresden- Radebeul under the supervision of Dr. med. Jost Kopp to prevent damaging the instrument by disassembling it or making incorrect measurements by gauging through the sound hole. In brief, CT imaging uses a series of x-ray measurements made from various angles, either by rotating the x-ray source (as was done here) or the specimen to be measured, to produce cross-sectional images of the object [6]. Through this process, cross-sectional images of the guitar body in all three dimensions have been created, with an interval of 1 mm between each image. To develop the geometric model of the guitar, measurements from over 2000 pictures have been evaluated and used. CT Scanned images of the guitar can be seen in Figure 1.

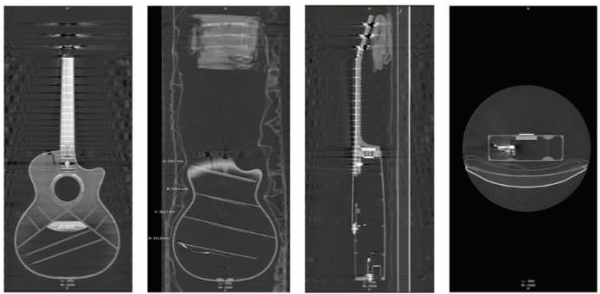

**Figure 1:** Different Cross-Sectional views of the CT Scanned Guitar.

The above images show the cross- sectional views that are parallel to the top and bottom of the guitar. The soundboard's strut width, bracing arrangement, and other visible components, such as the bridge's support and connection to the neck, can all be determined. The sound of the guitar is greatly influenced by the bracing pattern on the soundboard and back of the instrument, which also has a significant impact on vibration behavior. The bracing has two purposes: first, it strengthens the soundboard so that it can with-stand the forces generated by the strings; second, it gives control over the instrument's natural frequencies and mode shapes.

Although, modeling the guitar from the obtained CT Scanned images is not that easier, it gives a possibility to get the precise measurements. For viewing the images and to take measurements from them, Digital Imaging and Communications in Medicine (DICOM) viewer or software was used. The following is the image that shows the user interface of the DICOM viewer.

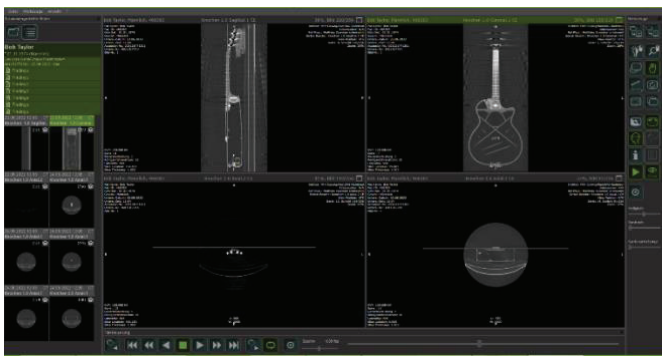

**Figure 2:** User interface of the DICOM viewer.

From the above image it can be seen that it is divided into four sections. Starting with top right section, it is the top view of the guitar, moving into counter clock vise direction, the second section shows the side view, third and fourth shows the axial or the side views of the guitar. The tool bar on the right side has the options to obtain the desired scaled measures. Thereby it became easier to extract required data from the CT Scanned images. Geometrical Model is designed using Autodesk Fusion 360. The following section summarizes the geometric model.

## Exploration of the Details and Geometrical Model

The guitar body is a complex geometry with different parts, some are directly visible and some are not. Below is the figure that shows some of the important parts of the guitar.

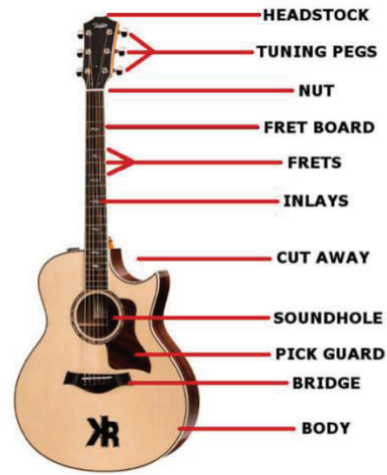

**Figure 3:** Components of the guitar [7].

The main body of the guitar is composed of the sound board, sound hole, side board, back board. These boards are made up of different orthotropic materials like Sitka Spruce and Tropical Mahogany. The bridge is the part of the guitar that holds the strings together near the sound board. A saddle is placed over the bridge and this minimizes the loss of vibrations. As already mentioned, pick guard prevents the guitar from being scratched by nails of the guitarist. There is a cut away and this makes the guitarist easier to reach the lower frets of the guitar. Head stock is the topmost part of the guitar and it holds the strings together on the top with tuning pegs.

The internal geometry of the guitar has braces, five set of braces on the back of the sound board and others on the back board of the guitar. Top braces provide sufficient strength to support the tension of the strings. It strengthens the top to prevent the neck from warping or part of the body from being lifted up when the strings are pulled. The other purpose of the braces is to create a certain timbre. Top harmonic braces create sharper sounds, while lower ones create softer sounds. Therefore, the braces can be considered as critical parts for an acoustical guitar and should be precisely modeled.

The design started with constructing the main body of the guitar. It includes the sound board on the top, the sound hole, sides and the back board of the guitar. While designing the back of the guitar there were some challenges, because its surface was not parallel. Thereby proper care was taken while designing the back. The below figure shows the back of the guitar.

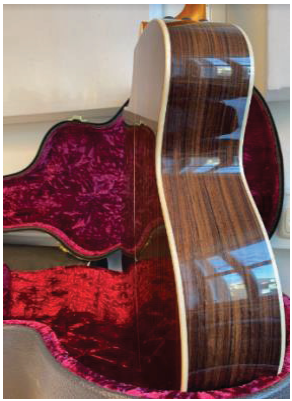

**Figure 4:** Back board of the guitar.

It can be seen that the back surface has curvature towards the sides along the axial line and also from top to bottom. To measure the thickness of the boards, Vernier calipers was used. Also the design was constructed using different planes in the software.

Later on the neck of the guitar was designed, with fret board on it. Even this was a challenging task since it has several curvatures and the geometry was existed in various planes. The Figure 5 shows the non- uniformity geometry of the neck.

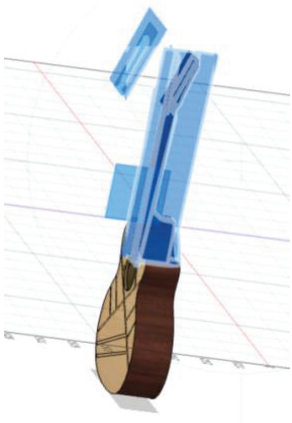

**Figure 5:** Neck of the guitar

The neck of the guitar was designed along with the fret board, head stock on the top and with tunings pegs. The neck is aligning in one plane and the head stock in the other plane.

The bridge of the guitar was modelled in different planes with saddle on top of it and plugs for the strings. The below figure shows the bridge with varying cross section and also in different planes.

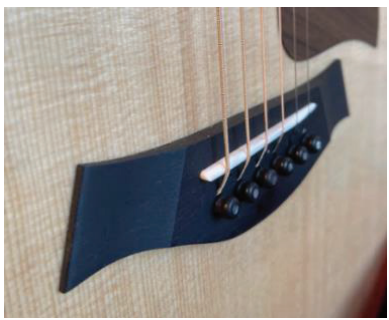

**Figure 6:** Bridge of the guitar.

Designing the internal parts of the guitar was a difficult task although there were the CT Scanned Images. The bracings have varying geometry both in height and thickness along its length and care was taken while modeling these. Below figure shows the varying geometry of the bracing. Location or positioning of the bracing is quite easier to design.

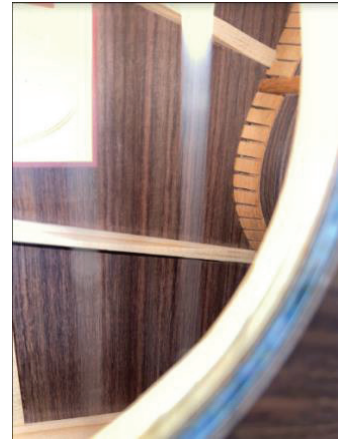

**Figure 7:** Bracing with varying geometry.

It was challenging to study the black and white CT Scanned images to detect some parts of the guitar. For example, heels within the guitar. Heels are the connecting junctions of both the main body and the neck of the guitar. There are some small blocks that binds the sides of the guitar with the boards, these can be seen in the Figure 7. These were also designed in addition to other parts like studs, tuning pegs and strings. The model contains more than 100 individual parts that have been designed in total. Below are the figures of both the original guitar and the CAD model. Figure 8 shows the top view of the whole guitar.

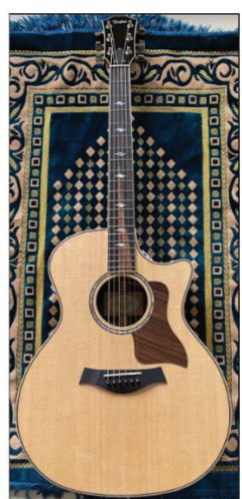 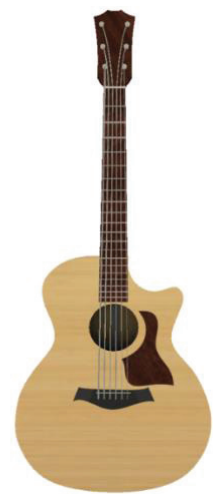

(a) Original Guitar    (b) CAD Model

**Figure 8:** Front view of both original guitar and the CAD model.

Efforts were made to design the model with minimal errors. The side view of the guitar is obtained and is represented in the figure below. Material selection for the visual aspects was included in the design. Color and the pattern were also considered.

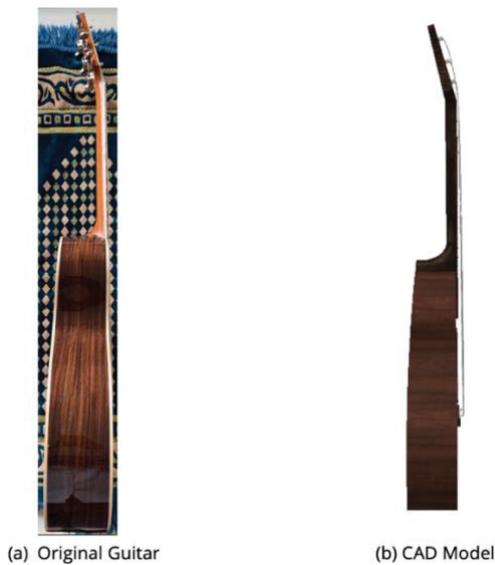

(a) Original Guitar     (b) CAD Model

**Figure 9:** Side view of both the original model and the CAD model of guitar.

Below figure shows the isometric view, cross- sectional view of the bottom of the guitar and the cross- section of the top plate of the guitar respectively.

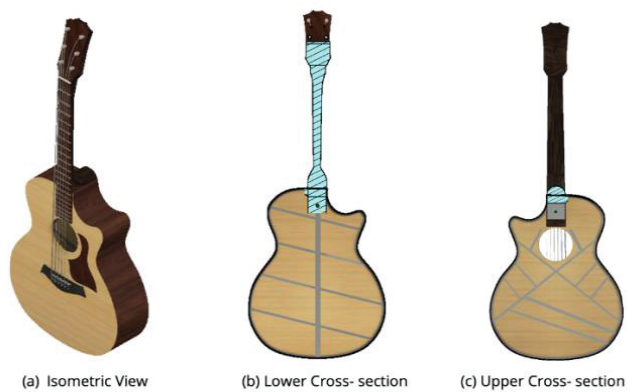

(a) Isometric View     (b) Lower Cross- section     (c) Upper Cross- section

**Figure 10:** Views of the CAD model.

The heels of the guitar can be seen in both the images of the cross-section of the guitar. They are located on the top center position of the sound board and below the heels. The bottom bracing is also shown in the second image and the third image shows the top bracing of the guitar.

## Conclusion

The aim of this study was to obtain a detailed 3D model of the guitar. There were quite few challenges during measurements and were tackled. Although, CT scan images produce very useful information about the structure, they are not sufficient to create model directly. For small details, a careful exploration on the instrument is required. In further studies, the obtained 3D model will be imported in a Finite Element Analysis software and a vibro- acoustic simulation model of the guitar will be developed.

## References

[1] Tahvanainen, Henna & Matsuda, Hideto & Shinoda, Ryo. (2019). Numerical simulation of the acoustic guitar for virtual prototyping, Conference: International Symposium of Musical Acoustics At: Detmold, Germany.

[2] URL: https://www.ancient.eu/Kithara

[3] URL: https://www.mi.edu/education/guitar-history-how-the-guitar-has-evolved/

[4] Evolution of the vibrational behavior of a guitar soundboard along successive construction phases by means of the modal analysis technique M.J. Elejabarrieta and C.S.A. Ezcurra, J. Acoust. Soc. Am., 108 (July) (2000), pp. 369-378

[5] https://www.taylorguitars.com/guitars/acoustic/814ce

[6] An entirely reverse-engineered finite element model of a classical guitar in comparison with experimental data, The Journal of the Acoustical Society of America 149, 4450 (2021); https://doi.org/10.1121/10.0005310, Alexander Brauchler, Pascal Ziegler, and Peter Eberhard

[7] URL: https://killerrig.com/guitar-part-names-and-what-they-do/